\documentclass[superscriptaddress,amssymb,twocolumn,aps]{revtex4}

\usepackage{graphicx,dcolumn,bm,amsmath,amssymb,color,bm}

\usepackage[utf8]{inputenc}
\usepackage[T1]{fontenc}
\usepackage{mathptmx}
\usepackage{gensymb}
\usepackage{soul,xcolor}
\setstcolor{red}
\usepackage[normalem]{ulem}  
\usepackage{xcolor}          

\newcommand{\JM}[1]{\textcolor{black}{#1}}
\newcommand{\JMone}[1]{\textcolor{black}{#1}}

\usepackage{hyperref}
\usepackage{color}
\usepackage{bm} 

\newcommand{\uvect}[1]{\hat{\bm{#1}}}
\newcommand{\vect}[1]{\bm{#1}}
\newcommand{\arowvect}[1]{\vec{\bm{#1}}}

\newcommand{\cuttext}[1]{%
  \ifmmode
    \text{\textcolor{red}{\sout{\ensuremath{#1}}}}%
  \else
    \textcolor{red}{\sout{#1}}%
  \fi
}

\newcommand{\fig}[1]{Fig.~\ref{#1}}

\definecolor{amber}{rgb}{1.0, 0.75, 0.0}

\newcommand{\eeq}{ \end{equation} }
\newcommand{\beq}{ \begin{equation} }

\newcommand{\eea}{ \end{align} }
\newcommand{\bea}{ \begin{align} }

\usepackage{xr}
\makeatletter
\newcommand*{\addFileDependency}[1]{
  \typeout{(#1)}
  \@addtofilelist{#1}
  \IfFileExists{#1}{}{\typeout{No file #1.}}
}
\makeatother
\newcommand*{\myexternaldocument}[1]{%
    \externaldocument{#1}%
    \addFileDependency{#1.tex}%
    \addFileDependency{#1.aux}%
}
\myexternaldocument{supplementary}

\begin{document}

\preprint{AIP/123-QED}


\title{Phase behaviour and defect structure of soft rods on a sphere}

\author{Jaydeep Mandal}
\affiliation{%
Centre for Condensed Matter Theory, Department of Physics, Indian Institute of Science, Bengaluru 560012, India
}%

\author{Hartmut Löwen}
\affiliation{Institut f\"ur Theoretische Physik II: Weiche Materie, Heinrich-Heine-Universit\"at D\"usseldorf, 40225 D\"usseldorf, Germany}

\author{Prabal K. Maiti}
\email{maiti@iisc.ac.in}
\affiliation{%
Centre for Condensed Matter Theory, Department of Physics, Indian Institute of Science, Bengaluru 560012, India
}%

\begin{abstract}

 Using particle-resolved molecular-dynamics simulations, we compute the phase diagram for soft repulsive spherocylinders confined on the surface of a sphere. While crystal (K), smectic (Sm), and isotropic (I) phases exhibit a stability region for any aspect ratio of the spherocylinders, a nematic phase emerges only beyond a critical aspect ratio lying between 6.0 and 7.0. As required by the topology of the confining sphere, the ordered phases exhibit a total orientational defect charge of \(+2\). In detail, the crystal and smectic phases exhibit two \(+1\) defects at the poles, whereas the nematic phase features four \(+1/2\) defects which are connected along a great circle. For aspect ratios above the critical value, lowering the packing fraction drives a sequence of transitions: the crystal melts into a smectic phase, which then transforms into a nematic through the splitting of the \(+1\) defects into pairs of \(+1/2\) defects that progressively move apart, thereby increasing their angular separation. Eventually, at very low densities, orientational fluctuations stabilize an isotropic phase. Our simulations data can be experimentally verified in Pickering emulsions and are relevant to understand the morphogenesis in epithelial tissues.
\end{abstract}

\maketitle


\section{\label{sec:intro}Introduction:}

Liquid crystals (LCs) exhibit mesophases which are characterized by a symmetry intermediate between that of isotropic fluids and crystalline solids \cite{de1993physics,onsager1949effects}. This unique coexistence of order and disorder renders them a subject of significant fundamental interest. Concurrently, LCs are exploited in numerous applications, prominently in optical systems and display technologies \cite{light1,light2,ferro1}, and increasingly in various emerging \JM{research} fields \cite{woltman2007liquid,lagerwall2012new,biosens1,van2000nematic}. Furthermore, the behavior of LCs under spatial confinement reveals a wealth of novel phenomena \cite{kim2016controlling,tortora2010self,mandal2025freezing,mandal2025melting,rajendra2023packing}. 
In particular, when constrained to spherical geometries, the interplay between topology and order leads to characteristic defect structures that disrupt the uniform orientational arrangement of the liquid-crystalline phase. On spherical substrates, the geometric constraint frustrates the local hexagonal order of a crystalline phase, enforcing a total defect charge of \(+12\). Such topological requirements are accommodated through twelve five-fold disclinations of unit charge, arranged in a triangulated pattern reminiscent of the stitching on a football \cite{bowick2009two}.

These topological constraints are not limited to crystalline order; in nematic liquid crystals, the reduced orientational symmetry gives rise to its own characteristic set of defect patterns \cite{dzubiella2000topological}. According to the Poincaré-Hopf theorem, the total charge of topological defects for a nematic on a sphere is \(+2\) \cite{poincare1885courbes}, which is typically realised by the existence of four \(+1/2\) defects on the sphere \cite{lubensky, vitelli2006nematic}. Interestingly, the location of the defect points is governed by the competition between the different elastic constants in the system. The Frank-Oseen elastic free energy density for a nematic is given by \cite{de1993physics,lubensky1998topological}:

\begin{equation}
    f = \frac{1}{2}[K_1 (\vect{\nabla} \cdot \vect{n} )^2 + K_2(\vect{n} \cdot (\vect{\nabla} \times \vect{n}))^2 + K_3(\vect{n} \times \vect{\nabla} \times \vect{n})^2]
\end{equation}
where \(\vect{n}\) denotes the director for the nematic phase, and \(K_1, K_2, K_3\) denote the splay, twist and bend elastic stiffness, respectively. For nematic shells of finite thickness, the defect configurations are influenced by the shell geometry and thickness \cite{vitelli2006nematic}. In the thin-shell limit on a spherical surface, however, the contribution from twist elasticity \(K_2\) becomes negligible. Previous theoretical and simulation studies \cite{shin2008topological,bates2008nematic} have demonstrated that, under the assumption of equal elastic constants, i.e. \(K = 1\), where \(K\) is the elastic anisotropy in the system defined as \(K = K_3/K_1\), the defect configuration adopts a tetrahedral arrangement that maximizes the separation between defect points. 
\JM{In the asymptotic limits of \(K \rightarrow \infty\) and \(K \rightarrow 0\), the splay and bend configurations minimize the free energy respectively and the resulting defect arrangements at the poles \cite{shin2008topological,allahyarov2017smectic} exhibit distinct angular profiles.}
The value of the elastic anisotropy can be controlled by physical parameters in a system, such as temperature or density \cite{dhakal2012nematic}. Remarkably, these theoretical predictions were confirmed through experiments \cite{lopez2011frustrated,lopez2011nematic,liang2011nematic} on liquid-crystalline shells, offering compelling evidence of the predicted defect structures.

Extending these investigations from nematic to smectic order, studies show that the resulting configurations depend on both the shell thickness and the anchoring conditions \cite{manyuhina2015thick,allahyarov2017smectic,wittmann2023colloidal,sharma2024smectic,jull2024curvature}. They shed light on the effects of spherical confinement on individual liquid-crystal phases, often through theoretical and computational approaches employing shape-anisotropic particle models \cite{bag2015molecular,lansac2003phase, maiti2002induced, maiti2004entropy, maiti2009computer}. Nevertheless, a comprehensive phase diagram for such systems confined to spherical geometry is still lacking. In this context, it is worth noting that in three-dimensional bulk, hard spherocylinder systems exhibit a rich phase diagram, encompassing isotropic, nematic, smectic, and crystalline phases \cite{mcgrother1996re}.  For hard or soft spherocylinders, nematic and smectic phases appear beyond certain critical aspect ratios (the ratio of length to diameter of a spherocylinder) \cite{bolhuis_frenkel}, consistent with Onsager’s theory \cite{onsager1949effects,graf1999phase,graf2007cell,troppenz2015nematic,allen2000molecular}. In two dimensions, smectic order is absent, and nematic order emerges only for aspect ratios above \(7.0\) \cite{bates2000phase}. While these phase diagrams \cite{marechal2011phase} are well established in three- and two-dimensional geometries, it would be highly interesting to construct the phase diagram on a spherical surface, where the influence of topology on phase stability can be systematically explored.

Here we seek to address that interest and examine another key question: how do defect configurations evolve during phase transitions on spherical topology? Understanding these changes is crucial, given the diverse significance of defects across soft matter systems\cite{li2015defects}. For instance, colloids can be functionalized with defect structures to create directional bonds \cite{devries}between neighbours. These can then be exploited to create colloidal crystals with unusual open lattices (such as diamond structures) with interesting perspectives for photonic band-gap materials. Moreover orientational defects contribute to the various biological processes \cite{saw2017topological,doostmohammadi2016defect,doostmohammadi2018active} such as providing driving forces of morphogenesis in epithelial tissues as exemplified recently for \textit{Hydra} where the location of defects dictates the formation and location of different limbs in the embryo \cite{maroudas2021topological}. This provides motivation to explore the issue of orientational defect structures on curved surfaces from a statistical physics point of view.


Therefore, we investigate the phase diagram for a system of soft repulsive spherocylinders (SRS) on the surface of a sphere, and analyse the defect structures in detail, using molecular dynamics (MD) simulations. The particle and force-field models used in the simulations are given in the next section. The results are presented in Section \ref{sec:results}, followed by the conclusions in Section \ref{sec:conclusion}.

\section{Simulation Details} \label{sec:simulation_details}

\begin{figure}[]
    \centering
    \includegraphics[width=1.0\linewidth]{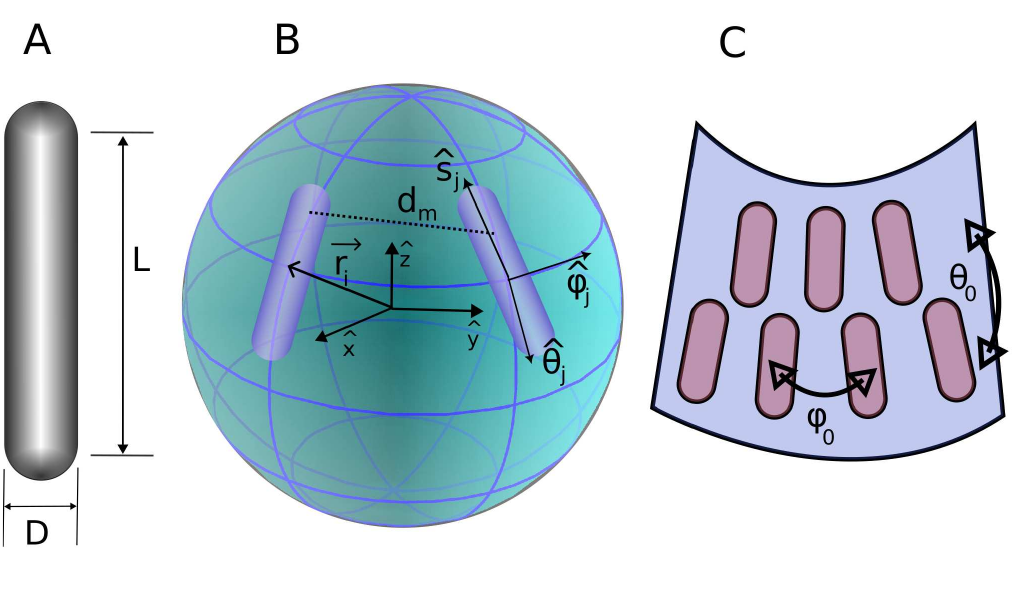}
    \caption{The schematics of the system: A) A soft repulsive spherocylinder (SRS) with body length \(L\) and diameter \(D\). B) The SRSs anchored tangentially on the surface of a sphere. The positions of the particles are denoted by the coordinates of their center of masses \(\arowvect{r_i}\), with the origin of the coordinate system at the centre of the sphere. The polar and azimuthal unit vectors \(\hat{\theta_j},\hat{\phi_j}\) are shown at the position of the \(j-th\) particle. \JM{The direction of the long axis of the \(j-th\) spherocylinder or the unit orientation vector is denoted as \(\uvect{s_j}\). The distance of closest approach between two spherocylinders is given by \(d_m\).} C) One particular face of the sphere between two latitudinal lines and two longitudinal lines is shown. For a layered smectic structure, the center of masses of the particles have an angular periodicity \(\theta_0\) along the longitude. For a crystal phase, an additional periodicity \(\phi_0\) appears along the latitude.
    }
    \label{fig:schematic}
\end{figure}

\begin{figure*}[]
    \centering
    \includegraphics[width=1.0\linewidth]{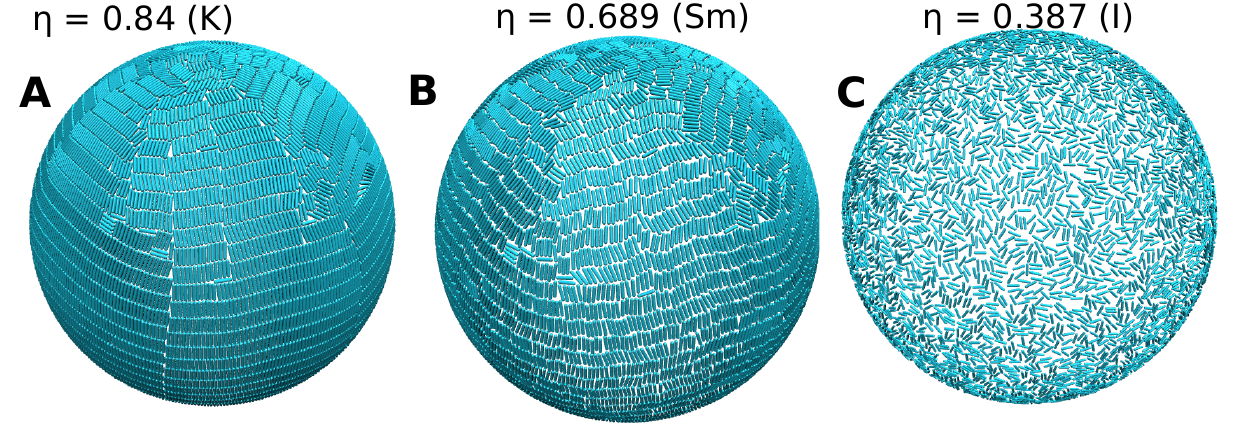}
    \caption{Snapshot of states obtained during expansion of a spherical surface of soft repulsive spherocylinders for the aspect ratio \(A = 4.0\), with \(N=10000\) at \(T^* = 1.0\) - A) The crystal phase with both orientational and positional ordering, obtained at a packing fraction of \(\eta = 0.84\). The orientations of the spherocylinders are ordered along the longitude of the sphere. The positional order is two-fold: it consists of multiple latitudinal layers that are stacked with inter-layer periodicity along the longitude, and it also features intra-layer periodicity along the latitude. B) At \(\eta = 0.689\), we observe a smectic phase which retains the layered structure with longitudinal periodicity, while the intra-layer ordering in latitudinal direction is lost. For both the states shown in A) and B), there is a \(+1\) defect at each of the poles. At a lower packing fraction of \(\eta = 0.387\), there is no positional or orientational ordering of the particles on the sphere (C). We observe no nematic phase in this particular value of the aspect ratio of the spherocylinders. The visualisations are done using VMD \cite{humphrey1996vmd}.
    }
    \label{fig:snapshots}
\end{figure*}

In this work, we simulate a system of soft repulsive spherocylinders anchored tangentially on the surface of a sphere. The length of the body axis of the spherocylinders is \(L\), with the diameter of the spherical part being \(D\) (Fig.\ \ref{fig:schematic}A). Note that the total body length of the spherocylinders is \(L+D\). The aspect ratio of the spherocylinders is given by \(A = L/D\).  An aspect ratio of \(0\) means a sphere of diameter D.

The center of masses (\(\arowvect{r_i}\)) of \(N = 10000\) spherocylinders are constrained on the surface of a sphere of radius \(R\) i.e. \(|\arowvect{r_i}| = R\), \JM{for all} \(i\in [1,N]\). The origin of the coordinate system is at the center of the sphere (Fig. \ref{fig:schematic}B). \JM{The direction of the long axis of the spherocylinders or the orientation vector of the particles are denoted by the unit vector \(\uvect{s_i}\). The tangential anchoring condition therefore imposes the restriction \(\uvect{s_i} \cdot \arowvect{r_i} = 0\)}. Strictly speaking, considering the diameter \(D\) of the spherocylinders, the system corresponds to a thin spherical shell of outer radius \(R+D/2\) and inner radius \(R-D/2\) in 3D.

The interactions between the SRSs are represented by a soft repulsive WCA pair potential \cite{weeks1971role},
\begin{align}
    U_{\rm WCA}(d_{m}) = \begin{cases}
            4\varepsilon\left[ \left( \frac{D}{d_m} \right)^{12} - \left( \frac{D}{d_m} \right)^6 \right] +\varepsilon, &d_m < 2^{1/6} D \\
            0 , \hspace{1cm} d_m\geq 2^{1/6}D
        \end{cases}
        \label{wca}
\end{align}
Here, $d_{m}$ denotes the distance of closest approach \cite{vega1994fast} between the spherocylinders (\fig{fig:schematic}B), and $\epsilon$ is the associated energy. We have used the reduced units throughout the manuscript, where energy and distances are scaled by \(\epsilon\) and \(D\), respectively. The reduced temperature \(T^*\) is computed by scaling the temperature value \(T\) with the factor \(\epsilon/k_B\), where \(k_B\) is the Boltzmann constant. We measured time in units of $D(m/\varepsilon)^{1/2}$, with \(m\) being the mass of the spherocylinder. 
The simulations are performed at \(T^* = 1.0\) in an NVT ensemble, maintained using a Berendsen thermostat \cite{berendsen1984bath} with a temperature coupling time of $\tau_T = 100\delta t$ where $\delta t = 0.001$ is the simulation timestep. The packing fraction \(\eta\) of the spherocylinders on the spherical surface is defined as \(\eta = \frac{NS_p}{\pi(2R+D)^2}\), where \(S_p\) is the area of projection of a spherocylinder on the spherical surface, which depends on the system parameters such as \(L,D\) and \(R\) \cite{allahyarov2017smectic}.

The simulation protocol has three steps -- in the first step, we prepare the initial configuration of the system at a high packing fraction (around \(\eta_i \approx 0.84\)). To achieve this, we first put \(N\) particles on the surface of a larger sphere of radius \(R_i = 2R_0\), where \(R_0\) is the required radius to obtain the desired value of packing fraction i.e. \(\eta = \eta_i\). The position of the center of masses of the particles in the initial configuration are obtained by following a Fibonacci sphere construction \cite{swinbank2006fibonacci} and the orientations of the particles are longitudinal (\(\uvect{s_i}(\arowvect{r_i}) = \hat{\theta_i}(\arowvect{r_i})\)), where \(\hat{\theta_i}(\arowvect{r_i})\) is the unit vector along the polar angle at  \(\arowvect{r_i}\) (Fig. \ref{fig:schematic}B). The whole system is then slowly deflated to a radius of \(R_f = R_0\) of the sphere. During deflation of the sphere, the positions and orientations of the particles were allowed to relax 
to avoid any jamming effect in the system.

In the second step, we run the molecular dynamics (MD) simulations of the system with the specified packing fraction \(\eta_i\) at \(T^* = 1.0\). The positions and velocities were updated using the velocity-Verlet algorithm, while the constraint equations 
\begin{equation}
    |\arowvect{r_i}| = R, \arowvect{r_i} \cdot \arowvect{v_i} = 0, \uvect{s_i} \cdot \arowvect{r_i} = 0 
\end{equation}
where  \(\arowvect{v_i}\) denotes the translational velocity of the \(i-th\) spherocylinder, were maintained using an adaptation of the RATTLE algorithm \cite{andersen1983rattle}. The orientations \(\uvect{s_i}\) of the SRSs obeys the equations of rotational motion \cite{mandal2025freezing} and they are updated in accordance with the rigid body dynamics of linear molecules \cite{rapaport2004art}. The simulations are run for a total time of \(t_N = 1.0 \times 10^6 \delta t\). This total duration was divided equally into two parts. The first half was allocated for the equilibration of the system, allowing thermodynamic quantities to reach stable equilibrium values. The remaining half \JMone{of the total simulation run-time} was designated as the "production phase," during which all data were recorded and subsequently analyzed. The final results shown in this manuscript are calculated using the trajectory from the production phase.

In the third step, we employed a stepwise expansion protocol to investigate the phase behavior as a function of packing fraction. Starting with the equilibrated high-density configuration (obtained from Step 2 above), the packing fraction \(\eta\) was reduced to a new target value by slowly expanding the sphere. The system was then re-equilibrated at this new packing fraction, and data were collected during the production runs. This expansion-equilibration-production cycle was iteratively repeated to generate configurations at progressively lower packing fractions. The packing fraction was systematically varied from an initial value of \(\eta_i \approx 0.84\) down to a final value of \(\eta_f = 0.1\) using a decrement of \(\Delta \eta \approx 0.035\) between successive runs. This step size consequently defines the minimum uncertainty in determining the packing fractions of any phase transitions. This entire simulation set, scanning all packing fractions, was conducted for SRS of several aspect ratios, namely \(A = 2.0,4.0,6.0,7.0,8.0\) and \(10.0\). Finally, to assess potential finite-size effects on the calculated phase-diagram, the complete set of simulations (across all \(\eta\) and \(A\)) was replicated for a larger system containing \(N = 20000\) particles.

An additional, important aspect of our methodology is that, as stated earlier, the initial structure has a longitudinal director configuration. High elastic anisotropy, corresponding to the asymptotic limit  \(K \rightarrow \infty\) can drive a system to adopt such configurations \cite{bates2008nematic,lopez2011nematic,dhakal2012nematic}. Conversely, we propose that, this initial condition of longitudinal orientation of particles effectively constrains the system to this high-anisotropy regime, implying a dominant bend elastic stiffness \(K_3 >> K_1\) in this system. Some of the resulting structures obtained from our simulations are consequently consistent with theoretical predictions for this limit. However, we have not explicitly calculated the elastic constants, as that task is non-trivial and can be accommodated in a future work.

After the simulations, we analysed the structures and identified the phases, based on the positional and orientational order in the system. The orientational order in the system is measured by the nematic order parameter \(S\), the largest eigenvalue of the traceless symmetric tensor $Q$ defined as \cite{de1993physics},
\begin{equation} \label{eqn:nematic-order-parameter}
    Q_{\alpha \beta} = \frac{1}{N}\sum\limits_{i=1}^{N} \frac{3}{2} s_{i \alpha} s_{i \beta} \ -\ \frac{1}{2}\delta_{\alpha \beta},
\end{equation}
where $i$ is the particle index while $\alpha,\beta$ corresponds to components of unit orientation vector $\uvect{s}$. 

The eigenvector $\arowvect{n}$ of \(Q\) corresponding to \(S\) denotes the director of the ordered phase. Due to the system's spherical topology, the global order parameter is an unreliable measure of the true nematic ordering. We therefore employed a local averaging method: the spherical surface was partitioned into a number of faces, the nematic order parameter was calculated for each, and these values were then averaged. This approach provides a more accurate depiction, as each small, quasi-planar face better represents the local ordering than the single global metric. 

We characterize the positional ordering of the particles' center of mass (COM) using two parameters: i)  \(\zeta\) \JMone{(as defined in eq \ref{equation smectic order parameter}),} for longitudinal ordering and ii) a radial distribution function in \(\phi\) (\(\phi\)-RDF) for latitudinal ordering. The longitudinal order parameter \(\zeta\) also serves as an indicator of the formation of multiple layers, as layers in this system typically 
\JM{span across the latitude and multiple such layers are stacked along the longitude. Therefore, considering the COMs of the particles within each layer forms a latitudinal line, which repeats itself along the longitude with the angular periodicity \(\theta_0\) (see Fig. \ref{fig:schematic})C.} 
The order parameter \(\zeta\) is designed to capture this specific periodicity.

\begin{equation} \label{equation smectic order parameter}
    \zeta = \frac{1}{N}\sum_i \exp[i 2\pi \frac{\Delta\theta_i}{\theta_0}]
\end{equation}
where \(\theta_i\) is the polar angle given by \(\theta_i = \cos^{-1}(z_i/R)\), \(z_i\) is the z-component of \(\arowvect{r_i}\). As noted above, this order parameter was also calculated by dividing the spherical surface into multiple sub-surfaces and then taking the average of the order parameter over them. Note that \(\Delta\theta_i\) refers to \(\theta_i - \theta_{min}\) where \(\theta_{min}\) is the minimum \(\theta\) value in each sub-surface. Additional azimuthal periodicity (\(\phi_0\)) develops within each of the layers at higher densities, which is revealed by the RDF calculation, done as follows: we first divide the sphere surface into different subsurfaces and calculated the azimuthal angle \(\phi_i\) for the center of mass of each of the particles. Then we subtracted \(\phi_{min}\) from each of these values, where \(\phi_{min}\) indicates the minimum value of \(\phi_i\) in that particular subsurface. We calculated a radial distribution function of the resulting values \(\Delta\phi\), and then took an average over different faces.

Using the order parameters described above, we analyzed the system's configurations to identify the phases for various SRS aspect ratios. We also investigated the role of defect structures within the ordered states during phase transitions. Strictly speaking, a "phase" indicates structures at the thermodynamic limit, whereas any confined surface will be inherently finite-sized. Therefore, the term "phase" in this context, as used throughout this manuscript, is to be understood as the finite-size steady state structures on the sphere.

\section{Results} \label{sec:results}

\subsection{Phases and Order parameters:} The common phases observed for all the aspect ratios of the SRSs considered in this work are crystal (K), smectic (Sm) and the isotropic (I) phase. In the crystal (K) phase, we observe that multiple layers are stacked along the longitude, and each layer spans across the latitude. The crystal (K) phase is thus defined by the presence of both the orientational and positional order in the system (Fig. \ \ref{fig:snapshots})A. The orientational order is given by the nematic director field along the longitudes. The positional ordering for a crystal phase is long-ranged in both the longitude \(\hat{\theta}\) and latitude \(\hat{\phi}\) directions. The longitudinal order arises from inter-layer periodicity, while the latitudinal order stems from the azimuthal periodicity within each layer (Fig. \ref{fig:snapshots}A).

The smectic (Sm) phase (see Fig. \ref{fig:snapshots}B) also possesses an orientational ordering and a positional ordering with layered structures. The smectic phase is distinguished from the crystal phase by its lack of intra-layer positional ordering of the particle COMs along the latitude (Fig. \ref{fig:snapshots}B). The isotropic phase lacks any orientational or positional order in the system (Fig. \ref{fig:snapshots}C). The nematic phase, which is observed only for the aspect ratios \(A > A_c\)\JMone{(the value of which lies between \(6.0\) and \(7.0\))}, shows an orientational ordering, (snapshot shown in Fig. \ref{fig:L8_OP}) but no positional ordering. 

The structural characteristics of the phases are identified by their respective order parameters. 
\JM{The \textit{crystal} phase exhibits high values of \(S\) and \(\zeta\) , and periodic peaks in the \(\phi\)-RDF. It is to be noted that, due to curvature, we never obtain the idealistic highest value of the order parameter as \(1\). The \textit{smectic} phase also shows similar characteristics in \(S\) and \(\zeta\) but lacks \(\phi\)-RDF peaks, as it has no intra-layer azimuthal periodicity. The \textit{nematic} phase, lacking a layered structure, is characterized by a high value of \(S\), but a low value of \(\zeta\). Finally, the disordered \textit{isotropic} phase shows low values for both \(S\) and \(\zeta\).}

\begin{figure}[]
    \centering
    \includegraphics[width=1.0\linewidth]{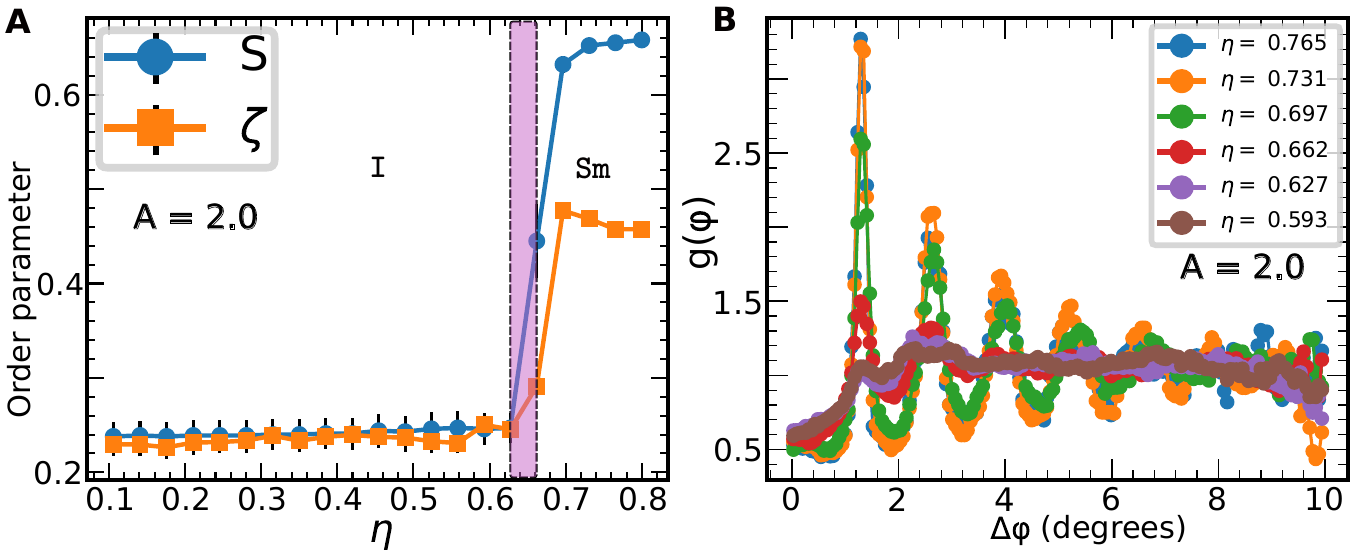}
    \caption{Determination of transitional packing fractions at phase transitions for \(A = 2.0\) for a spherical shell of soft repulsive spherocylinders. A) The nematic (\(S\)) and smectic (\(\zeta\)) order parameters plotted as a function of packing fraction \(\eta\). The nematic order parameter indicates the orientational ordering in the system, whereas the smectic order parameter indicates the presence of a layered structure. Both order parameters show a sharp rise at a packing fraction \(\eta_I \approx 0.627\), which demonstrates that the orientational ordering in the system appears with the layered structure and thus there is no nematic phase for the system. B) The radial distribution function of the \(\phi\) co-ordinates of the particles, plotted at different packing fractions. The appearance of a larger number of peaks in the plot at higher \(\eta\) shows the emergence of periodicity in the latitude direction, indicating a crystal-smectic transition.
    }
    \label{fig:L2_OP}
\end{figure}

\subsection{Phase behaviour:}

 \textit{Phase behaviour for \(A = 2.0\)}: For SRSs with an aspect ratio of \(A = 2.0\), decreasing the packing fraction by expanding the spherical surface (as described in Sec.~\ref{sec:simulation_details}) induces a sequential phase transition from a crystal(K), to a smectic(Sm), and finally to an isotropic(I) phase. In order to identify the phase boundaries, we calculated the nematic (\(S\)) (see eqn \ref{eqn:nematic-order-parameter}) and smectic (\(\zeta\)) order parameters (see eq \ref{equation smectic order parameter}) and plotted it as a function of \(\eta\), as shown in Fig. \ref{fig:L2_OP}A). The sharp rise in the values of the nematic order parameter indicates an orientational ordering transition, whereas the same for the smectic order parameter reflects the layering transition. Fig \ref{fig:L2_OP}A) shows that both transitions occur at the same packing fraction \(\eta = 0.627\), demonstrating that the orientationally ordered state has a layered structure. Therefore, we conclude that there is no nematic phase in the system. At and below this packing fraction \(\eta_{I2} = 0.627\), both \(S\) and \(\zeta\) have values close to \(0\), indicating an isotropic phase. Similarly, at and above the packing fraction \(\eta_{Sm1} = 0.662\), both attain a high value, exhibiting a smectic phase. Therefore, 
 \JM{we report that the packing fraction for the Sm-I transition is \(\eta_{Sm-I}= 0.64 \pm 0.02\), where the transition value indicates the average of the upper and lower limits, \(\eta_{Sm1}\) and \(\eta_{I2}\) respectively, and the uncertainly value stems from their difference, which appears due to the } discretised packing fraction interval of \(\Delta \eta\) used in our simulations, as we already mentioned in section \ref{sec:simulation_details}

 At higher packing fractions, the system shows signs of intra-layer periodicity, as captured by the emergent peaks in \(\phi\)-RDF plot shown in Fig. \ref{fig:L2_OP}B. We observe that multiple peaks begin to appear for the system between \(\eta_{Sm2} = 0.662\) and \(\eta_{K1} = 0.697\), reflecting the emergence of periodicity in the latitude direction, indicating a crystal phase. Therefore, 
 we conclude that the transition from crystal(K) to smectic(Sm) occurs within \JM{\(\eta_{K-Sm} = 0.68\pm0.02\)} for small rods with aspect ratio \(A = 2.0\).

\textit{Phase behaviour for \(A = 4.0\) and \(A = 6.0\)}:
We observe that, similar to the case of \(A = 2.0\), there is a simultaneous rise in both the values of \(S\) and \(\zeta\) for \(A = 4.0\) with \(\eta_{I2} =0.576\) and \(\eta_{Sm1} = 0.614\), indicating no nematic phase. Therefore, similarly we concluded that for the case of \(A = 4.0\), \JM{\(\eta_{Sm-I} =0.595\pm 0.02\)}. The melting transition occurs at \JM{\(\eta_{K-Sm} = 0.67 \pm 0.02\)}.

For \(A = 6.0\), we again confirmed the absence of the nematic phase. However, near the Sm-I transition at \(\eta_{Sm-I} = 0.57\), we identified a two-phase coexistence region (Sm-I) characterized by mutually uncorrelated layered structures coexisting with an isotropic fluid.

\begin{figure}[]
    \centering
    \includegraphics[width=1.0\linewidth]{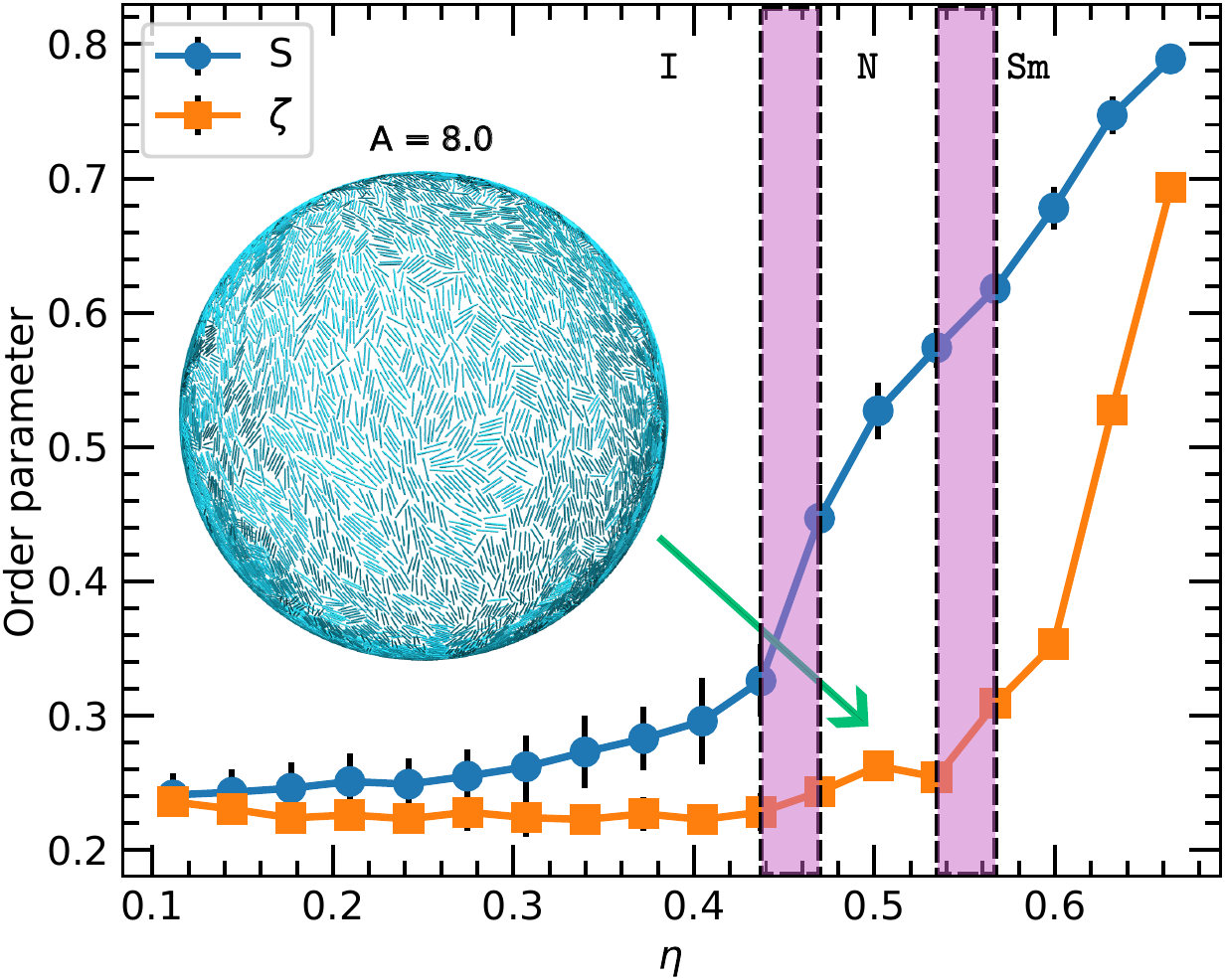}
    \caption{The existence of nematic phase for \(A = 8.0\). In the range \(0.470 \leq \eta \leq 0.535\), the structure shows a high value of \(S\), but a low value of \(\zeta\), indicating a nematic phase. The snapshot of the molecular configuration for a nematic phase at a packing fraction \(\eta = 0.5\) is also shown. \(\eta \leq 0.437\) shows an isotropic phase and \(\eta \geq 0.567\) exhibits smectic phase. The pink shaded areas denote the transition packing fractions.
    }
    \label{fig:L8_OP}
\end{figure}

\textit{Phase behaviour for \(A = 8.0\):} The clear evidence of the existence of the nematic phase can be observed in the system of \(A = 8.0\). In Fig. \ref{fig:L8_OP}), we can see three distinct regions: i) \(\eta \leq \eta_{I2}=0.437\) with both \(S\) and \(\zeta\) close to \(0\), indicating an isotropic phase. ii) \(\eta_{N1}=0.470 \leq \eta \leq \eta_{N2}=0.535\), \(S\) obtains a larger value but \(\zeta\) remains close to \(0\), demonstrating a phase with orientational ordering but no layered structure i.e. the nematic phase. In the nematic phase, the director field on the spherical surface is along the longitudinal directions as can be seen in the instanteneous snapshot shown in Fig. \ref{fig:L8_OP}. iii) \(\eta \geq \eta_{Sm1}=0.567\) showing high values for both \(S\) and \(\zeta\), indicating the smectic phase. The crystallization packing fraction was also observed to be close to \(0.7\), using the \(\phi\)-RDF analysis. Therefore, we finally arrive at the phase transition packing fractions given as: \JM{\(\eta_{K-Sm} =0.685 \pm 0.02, \eta_{Sm-N} =0.55 \pm 0.02, \eta_{N-I} =0.45 \pm 0.02\)}. 

Having confirmed the absence of a nematic phase for \(A = 6.0\) and its presence for \(A = 8.0\), we investigated the intermediate aspect ratio of \(A = 7.0\). Our analysis revealed that this system also exhibits a nematic phase, with the following transition packing fractions: \(\eta_{K-Sm} = 0.685 \pm 0.02\), \(\eta_{Sm-N} = 0.585 \pm 0.02\), and \(\eta_{N-I} = 0.485 \pm 0.02\).

For \(A = 10.0\), the system also exhibits the crystal (K), smectic (Sm), nematic (N) and isotropic (I) phases, and we determined the transition packing fractions using the same methodology as described before. Having identified the phase boundaries for all aspect ratios studied, we proceeded to construct the system's phase diagram.

\subsection{Phase diagram:}
\begin{figure}[]
    \centering
    \includegraphics[width=1.0\linewidth]{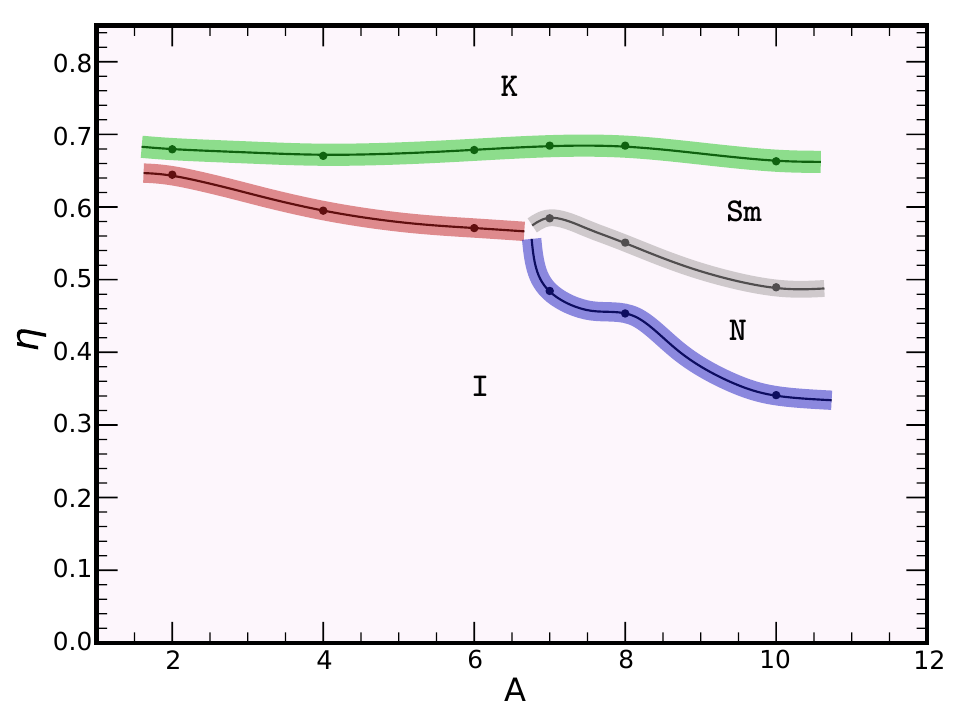}
    \caption{Phase diagram for a system of SRSs on the surface of a sphere at \(T^* = 1.0\). The crystal (K), smectic (Sm) and isotropic (I) phases are observed for all aspect ratios considered. The points indicate the packing fractions delimiting the phase transition. The black lines are the guides to the eye. The colored shaded regions represent the uncertainty in the transition packing fractions. The nematic phase (N) appears only above a critical aspect ratio of the SRSs (\(A_c\)), the value of which lies between 6 and 7, Above the critical aspect ratio \(A_c\), as the SRSs become longer, the transition packing fraction for the nematic phase decreases. The packing fraction for the K-Sm transition of the system is approximately around \(0.7\), which is similar to the freezing packing fraction for hard discs on a 2d plane \cite{li2022hard,bernard2011two,kapfer2015two} or sphere \cite{giarritta1992statistical}. 
    }
    \label{fig:phse_diagram}
\end{figure}

We show the phase diagram for a system of SRSs anchored tangentially on a sphere in Fig. \ref{fig:phse_diagram}. 
The black dots in the phase diagram indicate the transition packing fractions for the various phases. For example, at \(A=2.0\), the packing fraction values 0.68 and 0.64 correspond to the \textit{K-Sm} and \textit{Sm-I} transitions, respectively. The solid black lines connect these points to delineate the phase boundaries. Each black line is associated with a colored shaded region, which indicates the uncertainty in calculating the transition packing fraction. 
It is worth mentioning here that, we do not resolve a possible coexistence region in the phase diagram, as this is ill-defined for finite systems anyway. Our density estimates rather indicate region where a sharp cross-over between different equilibrium structures takes place.

The melting transition to the crystal phase (K) takes place at a packing fraction of \(\eta \approx 0.7\) for all the aspect ratios of the SRSs (see Fig. \ref{fig:phse_diagram}) studied in this work. The universality of the transition packing fraction agrees well with the earlier results reporting a similar freezing packing fraction for hard discs on a 2D plane or on a sphere \cite{huerta2006freezing,giarritta1992statistical}. Comparison with the 2D planar phase diagram (for hard spherocylinders\cite{bates2000phase}) highlights a key distinction: contrary to the planar system, the spherical surface \JM{does exhibit a smectic phase}. This Sm phase exists within specific packing fraction intervals that are dependent on the spherocylinder aspect ratio. As expected, the entropically-favored isotropic (I) phase (Fig.~\ref{fig:snapshots}C) is observed at low packing fractions for all aspect ratios.

In our simulations, we observe a nematic phase for \(A = 7.0\), but not for \(A = 6.0\), demonstrating that the nematics only appear for \(A > 6.0\). From the phase diagram, we observe that the lines delimiting the Sm-I, Sm-N and N-I transitions will converge within the limits \(A = 6\) and \(A = 7.0\), and therefore the \textit{critical aspect ratio} for the appearance of the nematic phase on the surface of a sphere is \(A_c \in (6,7)\). However, we presume that a greater number of detailed simulation with intermediate lengths of the SRS is required for a more accurate prediction of this I-N-Sm triple-point. Beyond this critical aspect ratio \((A > A_c)\), the packing fraction required to form a stable nematic phase decreases as the SRSs become longer (Fig.~\ref{fig:phse_diagram}). For comparison, we note that the critical aspect ratio for nematic on a 2D planar surface for hard spherocylinders is reported to be approximately \(  7.0\) \cite{bates2000phase}.

\begin{figure}[]
    \centering
    \includegraphics[width=1.0\linewidth]{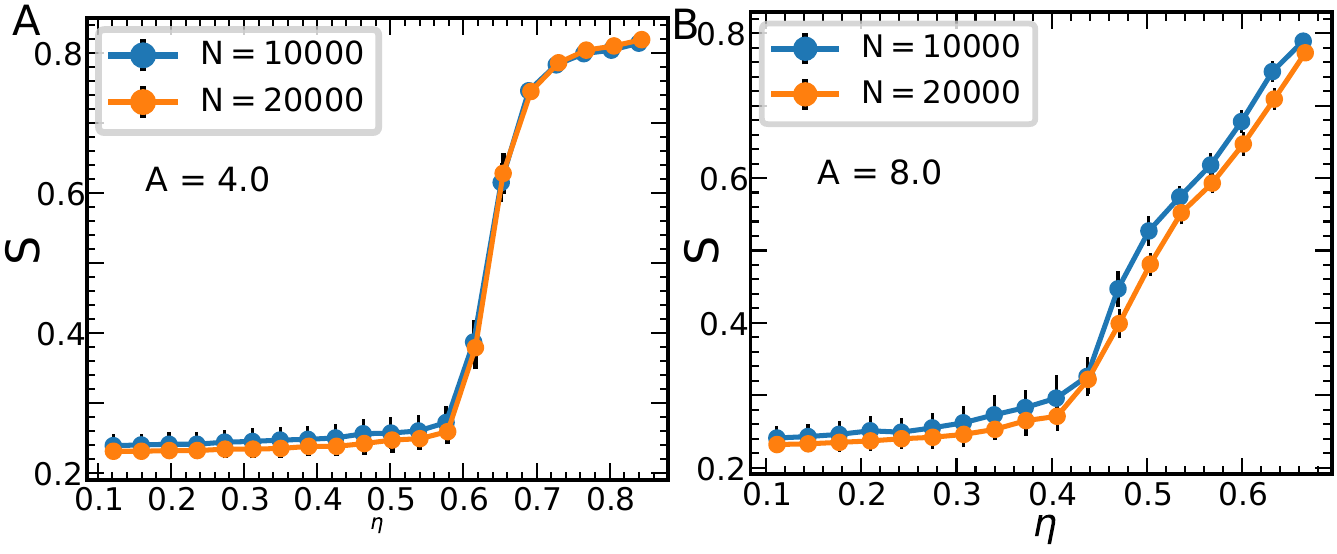}
    \caption{Effect of finite system size. Plot of the nematic order parameter \(S\) with packing fraction \(\eta\) for different system sizes \(N = 10000\) and \(20000\), shown for A) \(A = 4.0\), and B) \(A = 8.0\). For both aspect ratios, the nematic order parameter values remain almost independent of the system size. The transition gets a bit sharper for larger systems but the transition packing fraction itself is not significantly altered. 
    }
    \label{fig:system_size}
\end{figure}

\subsection{Effect of the system size:} The confined nature of the topological surface we are considering makes the system inherently finite-sized. Hence, it is important to check for the robustness of the phase diagram with the system size (or equivalently with the number of particles). Therefore, we also carried out simulations for  \(N = 20000\) SRSs. In Fig \ref{fig:system_size}A), we compare the nematic order parameters for \(N = 10000\) and \(N=20000\), for the aspect ratio \(A = 4.0\). The nematic order parameter is almost the same for both systems for all the packing fractions, but the profile is getting a bit sharper for large systems. This indicates first that the ordering transition and the transition packing fractions do not significantly alter with the system size and that there is a gradual shift towards the true thermodynamic bulk phase transition. Similarly, for \(A = 8.0\), the order parameters do not depend much on the system size, as shown in Fig. \ref{fig:system_size}B. Therefore, we can conclude that the phase diagram remains largely invariant with the system size. We should also note that, we might start seeing significant changes in the phase diagram and in the nature of the phase transitions if we consider very small values of \(N\), as the radius of the spherical surface will be small and the effect of surface curvature will begin to manifest.

\begin{figure*}[]
    \centering
    \includegraphics[width=1.0\linewidth]{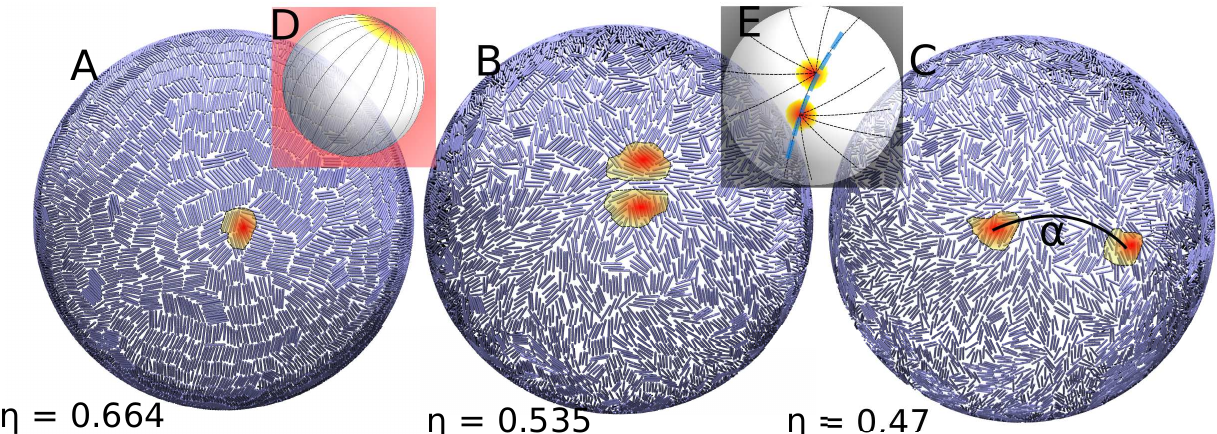}
    \caption{Defect structures at different phases and packing fractions. A) Snapshot of the smectic structure on the sphere at packing fraction \(\eta = 0.664, A = 8.0\). For simplicity, only the \(+1\) defect at the south pole is shown. The defect region is indicated by the yellowish-red region in the image. This defect at the south pole breaks in two \(+1/2\) defects when the system undergoes a phase transition to a nematic phase (B). Similar phenomena occur at the north pole, and therefore the total topological charge of the four defects in the nematic phase remains \(+2\). (C) At a lower packing fraction, within the nematic region, the angular separation between the two \(+1/2\) defects, denoted by \(\alpha\), increases. However, the line joining the four defects still maintains a great circle geometry. \JMone{Note that \(\alpha\) is defined by the angle between the position vectors of the two defect cores at any of the poles.} D) Schematics of the director configuration (indicated by the lines on the sphere) for a smectic with a \(+1\) defect at each pole, indicated by the yellowish-red region (north pole is visible in the image, south pole also possesses another defect of topological charge of \(+1\)). E) Top-view (from above the north pole) of the schematics of the director configuration on the sphere for a nematic phase. A similar structure can be theoretically obtained by following the "cut-and-rotate" procedure as mentioned by \cite{shin2008topological}. The defects lie on a great circle as indicated by the blue dashed line, and the angular separation between the defects \(\alpha\) is along this connecting line.
    }
    \label{fig:defect_phase_transition}
\end{figure*}

\subsection{Defect structures and ordering transition:}
Having established the independence of the phase diagram over the size of the system, we now consider into the structure and effect of the topological defects during the phase transitions.

As evident from the phase diagram, the aspect ratio of \(A = 8.0\) exhibits all the four phases (K, Sm, N, I) at different packing fractions. Therefore, we chose \(A = 8.0\) to study the defect structures for the various ordered phases. The crystal phase has a longitudinal director configuration with a defect of topological charge \(+1\) at each pole. When the packing fraction is reduced, the system shows a smectic phase with the nematic director still maintaining a longitudinal configuration. In Fig. \ref{fig:defect_phase_transition}A), we show a representative snapshot of the system in the smectic phase, near the south pole. There exists a defect of topological charge \(q_s = +1\) and angular phase \(\gamma_s = 0\) (\JMone{a measure of the angle between the radius vector to any point from the defect core and the nematic director at that point}, see \cite{allahyarov2017smectic}). Therefore, the total topological charge of the defects on the surface of the sphere is \(+2\). The schematics of the defect configuration is given in Fig. \ref{fig:defect_phase_transition}D. The defects are indicated by the yellowish-red region in the images. 

At an even lower packing fraction, each of the topological defect of charge \(+1\) at the poles splits into two defects with the topological charge \(q_s = +1/2\), and an angular phase of \(\gamma_s = 0\) (see Fig. \ref{fig:defect_phase_transition}B). The total topological charge of the nematic defects on the sphere is conserved (\(+2\)). The splitting of the defect charge marks the onset of the phase transition into the nematic phase. In Fig. \ref{fig:defect_phase_transition}B), we have shown the defect configuration on the south pole of the sphere. Considering all four defects on both the poles of the sphere, we observe that the defects arrange themselves in a great circle arrangement, which has been observed in previous studies as well \cite{bates2008nematic,dhakal2012nematic,shin2008topological,lopez2011nematic}.

As the packing fraction within the nematic phase decreases, the defect configuration still preserves the great circle geometry, but the angular separation between the defect structures \(\alpha\) increases (see Fig. \ref{fig:defect_phase_transition}C). The angular separation, \(\alpha\), is defined by the angle between the \JMone{position vectors of the two defect cores at the pole}. For the low-density nematic, the director configuration with the defects, near one of the poles, is shown in Fig. \ref{fig:defect_phase_transition}E). Such configurations are obtained by a "cut-and-rotate" surgery on a purely longitudinal director configuration with a \(+1\) topological defect charge at each pole, as has been previously pointed out by Ref.\cite{shin2008topological}. The great circle arrangement of the topological defects are shown in the blue dashed line in Fig. \ref{fig:defect_phase_transition}E. We observe that, when we decrease the packing fraction even further, the orientational fluctuation of the spherocylinders increase, and finally the system makes a phase transition in the disordered isotropic phase.

\JM{This might raise the question of the conditions required to observe a tetrahedral defect arrangement, which is known to arise in the limit \(K=1\) \cite{dzubiella2000topological,dhakal2012nematic}. In contrast, the structures in our simulations are consistent with the \(K \rightarrow \infty\) limit, as inferred from the system's initial ordering (see Sec.~\ref{sec:simulation_details}). Thus the tetrahedral defect arrangement might only occur for very long rods. Achieving the \(K=1\) limit necessitates alternate approaches; We can start with randomly distributed orientations of the particles and compress the system, but this methodology suffers from the serious concern of particles getting arrested due to jamming, which would prevent the system from reaching its true equilibrium phase. Another possible way might be thermal annealing of the system, which relies on exploiting the temperature dependence of the elastic constants, but the required quench temperature is not known \textit{a priori}. Either way, we believe the phase diagram will remain largely invariant.}


\section{Conclusion} \label{sec:conclusion}

In this work we have simulated a system of soft repulsive spherocylinders tangentially confined to the surface of a sphere at fixed temperature, exploring a wide range of packing fractions for different aspect ratios (\(A\)). Four stable phases are observed: i) Isotropic (I), with no ordering in the system, ii) Smectic (Sm)- with orientational ordering and layered structures along longitudes, iii) a crystal (K) phase with orientational ordering and positional ordering in longitude as well as latitudes, iv) a nematic phase (N). The latter, however, is only stable beyond a critical aspect ratio lying between 6.0 and 7.0. The system exhibits sharp crossovers between these phases which we quantify by the use of suitable order parameters. These associated transitional packing fractions were computed and found to be consistent with earlier data obtained for planar hard spherocylinders \cite{bates2000phase} and hard disks on the sphere
\cite{giarritta1992statistical}.

We have then localized the topological defects in the orientationally ordered states and linked their relation to the phase transitions. The crystal and smectic phases have a longitudinal director configuration with defects at each of the poles, with topological charge \(+1\). At a lower packing fraction, each of the \(+1\) defect splits into two \(+1/2\) defects, marking the onset of phase transition into the nematic phase. The four topological defects have a total charge of \(+2\) and arrange themselves in a great circle arrangement. Within the nematic phase, decreasing the packing fraction increases the angular separation between the defects, while still maintaining the great circle arrangement. Further decreasing the packing fraction destroys the orientational ordering in the system by increasing the fluctuations, and the system undergoes a phase transition into a disordered isotropic phase.


Directions of future work based on our finding are multifarious: First of all rod-like particles forming smectic phases on manifolds more complicated than a sphere should be studied \cite{alexander2012colloquium} even if there are cusps \cite{senyuk2013topological,wittmann2021particle,monderkamp2021topology,monderkamp2022topological}. It would be highly interesting to see how the topology of the manifold is encoded in the defect locations. Also more complex particle shape such as chiral ones \cite{monderkamp2023network,chattopadhyay2024stability} can be considered. Finally, while the current results concern equilibrium questions, the model can be extended towards non-equilibrium situations\cite{decamp2015orientational,paliwal2020role}. Examples include the inflation and deflation process of the sphere itself \cite{janssen2017aging} or particle self-propulsion, i.e. activity \cite{bechinger2016active,bowick2022symmetry,venkatareddy2023effect,venkatareddy2025growth}. Active particles on the sphere have been studied both on the particle-resolved level and within field-theoretical approaches for different phases \cite{sknepnek2015active,praetorius2018active,nestler2024active} and it would be interesting to see how the phase diagram obtained here will be changed when activity is present.

\begin{acknowledgments}

JM thanks MHRD, India for the fellowship. PKM thanks DST, India for financial support and SERB, India for funding and computational support. HL thanks the German Research Foundation (DFG) for funding under LO 418/29.

\end{acknowledgments}

\section*{Conflict of Interest}
The authors have no conflicts to declare.

\section*{Data Availaiblity}
The data that support the findings of this study are available from the corresponding author upon reasonable request.



\bibliography{ref}


\appendix

\renewcommand\thefigure{\thesection.\arabic{figure}}    
\
%
\end{document}